# Direct imaging of chiral domain walls and Néel-type skyrmionium in ferrimagnetic alloys

Boris Seng[1,2,3,4], Daniel Schönke[1], Javier Yeste[1], Robert M. Reeve[1,3], Nico Kerber[1,3,4], Daniel Lacour[2], Jean-Loïs Bello[2], Nicolas Bergeard[5], Fabian Kammerbauer[1], Mona Bhukta[1], Tom Ferté[5], Christine Boeglin[5], Florin Radu[6], Radu Abrudan[6], Torsten Kachel[6], Stéphane Mangin[2], Michel Hehn[2] and Mathias Kläui[1,3,4, †]

1. Institut für Physik, Johannes Gutenberg-Universität Mainz, Staudingerweg 7, 55128 Mainz, Germany

2. Institut Jean Lamour, UMR CNRS 7198, Université de Lorraine, 2 allée André Guinier, 54011 Nancy, France

3. Graduate School of Excellence Materials Science in Mainz, Staudingerweg 9, 55128 Mainz, Germany

4. Max Planck Graduate Center mit der Johannes Gutenberg-Universität, Staudingerweg 9, 55128 Mainz, Germany

5. Université de Strasbourg, CNRS, Institut de Physique et Chimie des Matériaux de Strasbourg, UMR 7504, F-67000 Strasbourg, France

6. Institut für Methoden und Instrumentierung der Forschung mit Synchrotronstrahlung Helmholtz-Zentrum Berlin für Materialien und Energie GmbH, Albert-Einstein-Str. 15, 12489 Berlin, Germany

† Corresponding author: klaeui@uni-mainz.de

## Abstract

The evolution of chiral spin structures is studied in ferrimagnet Ta/Ir/Fe/GdFeCo/Pt multilayers as a function of temperature using scanning electron microscopy with polarization analysis (SEMPA). The GdFeCo ferrimagnet exhibits pure right-hand Néel-type domain wall (DW) spin textures over a large temperature range. This indicates the presence of a negative Dzyaloshinskii-Moriya interaction (DMI) that can originate from both the top Fe/Pt and the Co/Pt interfaces. From measurements of the DW width, as well as complementary magnetic characterization, the exchange stiffness as a function of temperature is ascertained. The exchange stiffness is surprisingly mostly constant, which is explained by theoretical predictions. Beyond single skyrmions, we find by direct imaging a pure Néel-type skyrmionium, which due to the absence of a skyrmion Hall angle is a promising topological spin structure to enable high impact potential applications in the next generation of spintronic devices.

## Introduction

The competition between different interactions in magnetic materials leads to a wide variety of different magnetic spin structures from single-domain states to chiral spin textures. Among them, magnetic skyrmions in thin-film multilayer systems [1-7] are nowadays widely studied due to their attractive properties for potential applications including their room temperature (RT) stability [8-10]. Skyrmions are topological spin structures that can be stabilized through an antisymmetric exchange interaction that arises in a system with a broken inversion symmetry: the Dzyaloshinskii-Moriya interaction (DMI) [11-12]. Recent studies confirmed the current-driven dynamics of skyrmions in ultrathin ferromagnets via spin-orbit torques (SOTs) that act efficiently on Néel-type spin textures [1,8,13]. For deterministic dynamics, sufficient DMI is then required to yield homochiral domain walls (DW). Therefore, skyrmions stabilized with DMI are particularly suitable for next generation spintronics devices, such as the skyrmion-based racetrack memory [8]. However, the dynamics of ferromagnetic skyrmions via SOTs exhibits a transverse motion due to their non-zero topological charge [14-16]. This behavior is especially unwanted in many applications since skyrmions could be annihilated at an edge of a device with the information carried by the skyrmion lost as a result. Materials with antiferromagnetically exchange-coupled magnetic sublattices have been proposed to reduce or annihilate this skyrmion Hall effect owing to the overall zero topological charge [17-18].

Ferrimagnetic materials such as rare earth (RE) – transition metal (TM) alloys are made of two antiferromagnetically exchange-coupled magnetic sublattices. For a given stoichiometry, at a temperature called the magnetization compensation temperature, the magnetization of the two sublattices is equal and opposite and therefore the net magnetization is zero [19]. Similarly, an angular momentum compensation temperature can be defined where a zero skyrmion Hall angle is predicted [20]. GdFeCo ferrimagnetic alloys have attracted significant attention since the discovery of all-optical switching (AOS) in these alloys [21]. Since then two types of AOS has been observed [22-23] and the possibility to reverse the magnetization with a single electron pulse has been demonstrated [24]. In particular, to use these systems for topological spin structure devices, one needs deterministic behavior in devices and for that it is necessary that the DMI is sufficient to generate a single chirality of skyrmions with a Néel-type spin texture. This then enables efficient SOT driven DW or skyrmion motion. So to realize this, one requires a detailed analysis and high resolution magnetic imaging on the internal spin structures of such ferrimagnetic skyrmions, which is still missing to date. In particular, as ferrimagnet can exhibit strong temperature dependence of the properties, one needs to ascertain the temperature range where robust properties are found.

In this study, we investigate the chirality of spin textures in a ferrimagnet namely Ta/Ir/Fe/GdFeCo/Pt by imaging the spin structure of the domain walls using SEMPA [25-27]. This surface-sensitive imaging technique has already been successfully used to determine the chiral character of out-of-plane (OOP) magnetized spin textures in ferromagnetic materials [28-29] and here we demonstrate that we can determine the chiral character of spin textures also for ferrimagnetic materials. From the SEMPA images, we are also able to extract the domain wall width across a wide range of temperatures, which allows us to determine the exchange stiffness evolution with temperature as a crucial parameter that governs the stability and operation temperature range.

**Experimental details**

A multilayer thin film of Si//Ta(5)/Ir(5)/Fe(0.3)/Gd$_{26.1}$Fe$_{65.5}$Co$_{8.3}$(8)/Pt(5) is deposited using magnetron sputtering in a chamber with a base pressure of $2.4 \times 10^{-8}$ Torr, with the thickness of each individual layer given in nanometers in parentheses. The ferrimagnetic layer was grown by co-sputtering and the atomic compositions were estimated from the deposition rates of each target.

The $Gd_{26.1}Fe_{65.5}Co_{8.3}(8)$ alloy has been chosen for its low coercivity that makes the stabilization of OOP spin textures by magnetic fields easier. The CoFe dominant alloy phase has been selected to provide a lower compensation temperature and lower resistivity to avoid sample heating effects. Interfacial DMI comes from a strong spin-orbit coupling with broken inversion symmetry between a heavy metal (HM) and ferromagnetic materials (FM) that arises mainly from the first atomic layer [30]. An ultrathin Fe(0.3) has been inserted between Ir and the ferrimagnetic layer to enhance the interfacial DMI which is known to be particularly strong and positive at the Ir/Fe interface. However, we note that the DMI at the Ir/Fe interface has also been calculated to be negative [31]. At the other interface, the DMI at the Co/Pt or Fe/Pt interfaces is expected to be strong and negative [2,30,32].

**Results and discussion**

The sample exhibits perpendicular magnetic anisotropy (PMA) with a small switching field after milling (see Supplementary S1). The stabilization of spin textures is achieved by cycling an in-plane (IP) magnetic field at RT where structures start to nucleate randomly. Figures 1(a) and 1(b) show the direction of the in-plane magnetization, imaged with SEMPA (see Supplementary S2), under zero magnetic field after the nucleation process. Contrast in the DWs can be observed where the in-plane magnetization is expected with a brighter contrast present for the top (left) of each spin textures indicating an up (left) tendency for the local direction of the magnetization. Conversely dark contrast is seen at the bottom and right edges of the domains as shown in figures 1(a,b), the absolute in-plane magnetization image has been generated in figure 1(c) where the brighter contrast indicates the in-plane saturated magnetization. This contrast clearly indicates the position of the domain walls where in-plane components are then present. From the analysis, the position of the skeleton of the domain walls, i.e. the line at the center of the DWs, can then be defined. The determination of the skeleton allows us to establish the DW intensity profile shows in figure 1(d) by averaging the measured local DW profile at each position of the skeleton (see Supplementary S3). However, the imaged DW profile needs to be analyzed to obtain the true profile, principally due to the broadening of the features due to the finite size of the beam profile. The experimental profile can be approximated as the convolution of the theoretical DW profile with a Gaussian function describing the electron beam distribution as follows:

$$DW_{profile} \propto \, cosh^{-1}\left(\frac{x}{\Delta}\right) \otimes e^{-\frac{x^2}{2\sigma^2}} \ (1)$$

where the hyperbolic function represents the real DW profile with $\Delta$ the Bloch parameter. In our case, we use for the domain wall width $\delta$, defined by Lilley [33] where $\delta = \pi\Delta$. To determine the Gaussian function, we considered a sharp defect modeled as a step function (see Supplementary S4). The Gaussian is found to be narrower than the imaged DW and here it has a rather small but non-negligible influence on the domain wall width measurements. From this analysis, we finally determine the real DW width to be $\delta_{real} = 175 \pm 5 \, nm$ at RT.

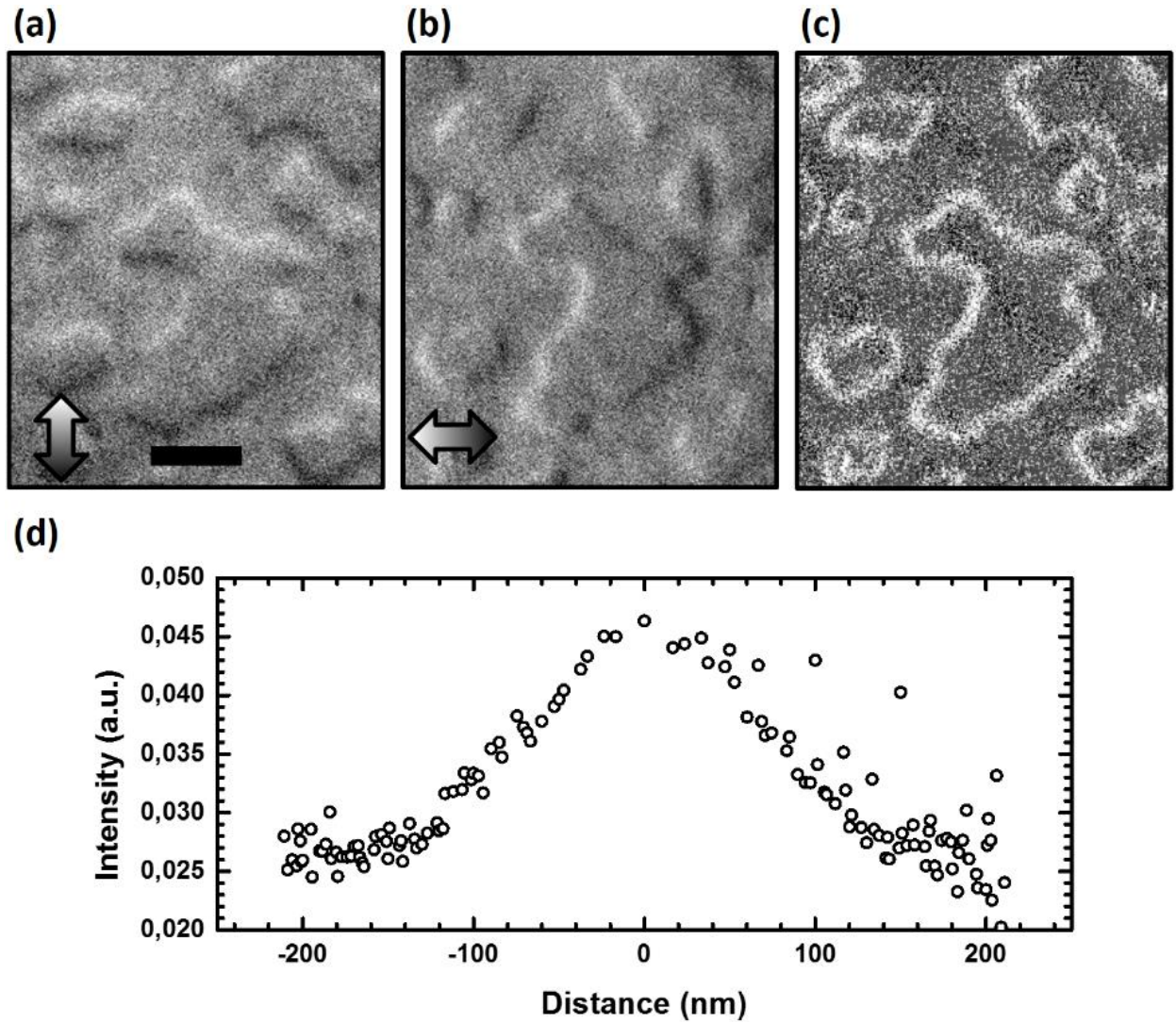

**Figure 1.** Determination of the average DW width of spin textures in Ta/Ir/Fe/GdFeCo/Pt taken with SEMPA at RT. (a) Vertical and (b) horizontal in-plane components of the magnetization. The direction of magnetization is indicated by the grayscale contrast as displayed on the double arrows. Scale bar in (a): 1 µm. (c) Reconstruction of the absolute in-plane magnetization intensity. The white contrast indicates a saturated in-plane magnetization. (d) Distribution of the average domain wall intensity profile for all the spin textures present in the image.

After having defined the DW width, we analyze the spin distribution within the DWs. Figure 2(a) displays the direction of the in-plane magnetization in the DWs by way of the color wheel shown in the inset. Qualitatively, we first see that all the magnetic structures present the same chirality, namely they are homochiral clockwise Néel-type spin textures. For a precise analysis, we then compare the spin direction of each position inside the DW with the direction of the local tangent of the DW (figure 2(b)) (see Supplementary S3). The histogram indicates that the magnetization direction in the DW structures forms a distribution centered around -90°. We conclude that our material presents a pure clockwise Néel-type homochiral character, which can be explained by the negative sign of the interfacial DMI in typical Fe/Pt and Co/Pt interface when the Pt layer is on top of the Fe or Co layer. Since SEMPA is a surface sensitive technique, we only probe the direction of the magnetization close to the surface. To ascertain that this chiral character is not a local effect at the surface due to flux closure but is indeed supported through the whole thickness of the film, asymmetric bubble expansion has been performed that confirms our clockwise Néel homochirality in the full film [34].

As explained above, an IP magnetic field is used to nucleate spin textures that start to propagate at random positions. Remarkably, using an IP oscillating magnetic field with a certain damping ratio allows for the nucleation of a spin texture inside another spin texture that has been nucleated previously. In figures 2(c) and (d), one can observe a small 160 nm diameter skyrmion stabilized in a larger skyrmion bubble, forming a magnetic skyrmionium, topological spin texture that is especially attractive due to its zero topological charge, which leads to a vanishing skyrmion Hall angle that is promising for application [35-36].

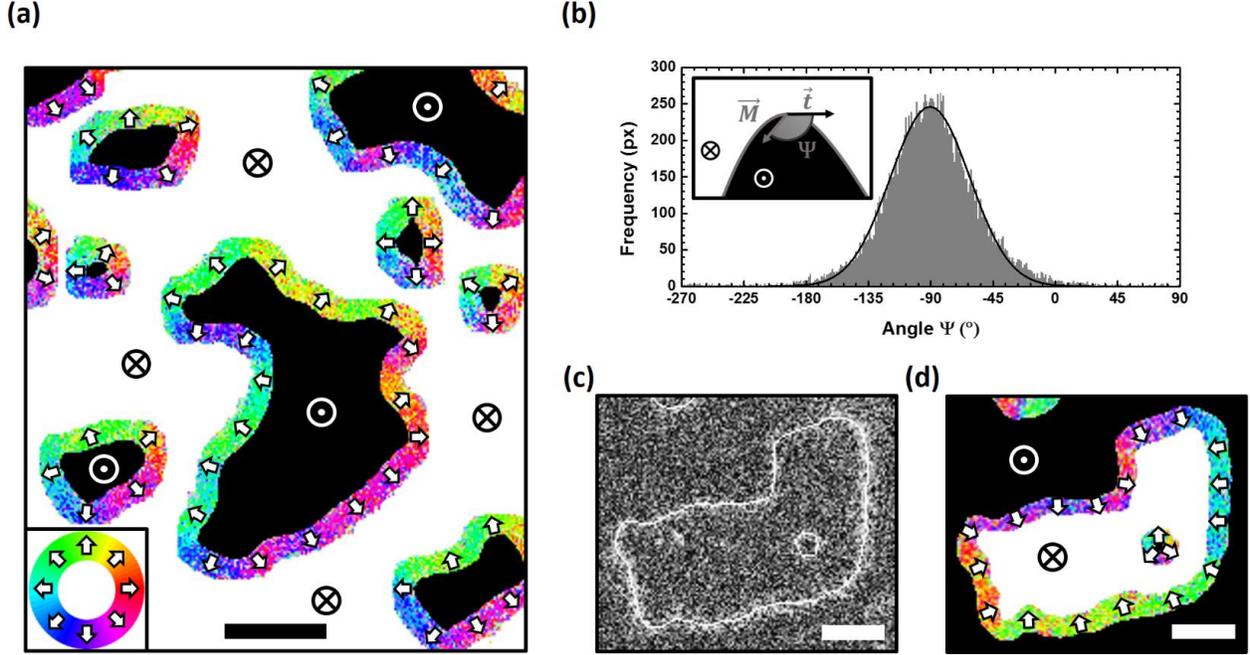

**Figure 2.** Determination of the magnetization direction inside the DWs of different ferrimagnetic chiral spin textures. (a) The direction of the in-plane magnetization in the domain walls is displayed as defined by the color wheel in the bottom left corner of the image. Scale bar in (a): 1 μm. (b) Distribution of the direction of the in-plane magnetization $\vec{M}$ in the domain wall with respect to the local tangent $\vec{t}$ at RT: angle $\psi$ (see inset). A Gaussian fit indicates a central value around -90°. (c) Absolute in-plane magnetization intensity with skeleton displayed in white and (d) direction of the in-plane magnetization in the domain walls of a ferrimagnetic skyrmionium at 320 K. Scale bars in (c) and (d): 500 nm.

To extract the keys magnetic parameters of the system, we analyze the measured domain wall profiles by considering the model put forward in I. Lemesh *et al.* [37]. In that work, the authors demonstrated that the DW width $\Delta$ is given by:

$$\Delta(d, \Psi) = \Delta_0 - \frac{1}{\frac{2\pi(Q-1)}{d} + \frac{1}{\Delta_0 - \Delta_\infty(\Psi)}} \quad (2)$$

where $d$ is the thickness of the magnetic material and $\Psi$ is the domain wall angle; $\Delta_0 = \sqrt{\frac{A}{K_{eff}}}$, $\Delta_\infty = \sqrt{\frac{A}{K_u + \frac{\mu_0 M_S^2}{2} sin(\Psi)^2}}$ , $Q = \frac{2K_u}{\mu_0 M_S^2}$, $K_{eff} = K_u - \frac{\mu_0 M_S^2}{2}$ depending on the exchange stiffness $A$, the uniaxial anisotropy $K_u$ and the saturation magnetization $M_s$.

The DMI dependence of the DW width enters via the angle $\Psi$. In the case of pure-Néel type domain walls, above a certain DMI threshold value, the exact value of the DMI does not affect the DW width since $|\Psi| = 90°$. In this case, by measuring the saturation magnetization $M_s$ and the effective anisotropy $K_{eff}$ (defined as the difference in the areas of the IP and OOP hysteresis loops) using a Superconducting Quantum Interference Device (SQUID), the exchange stiffness can be evaluated from the DW profile to be $A = 8.0 \pm 0.5 \, pJ.m^{-1}$ at RT. Therefore, through the measurement of the domain wall width via SEMPA, it is possible to determine the exchange stiffness of a material, a parameter that is not easily accessible using other simple technics.

One crucial point for the use of magnetic skyrmions in spintronic devices is that the homochiral character of the DWs is preserved over a large temperature range. Therefore, to assess this, the previous analysis has been carried out from 150 K to 315 K. We see in figure 3(a) that the pure Néel character remains for this temperature range. Next, we analyze the DW width for these temperatures where the different values are reported in figure 3(b). We find that the DWs are narrower when the temperature is decreased due to the increase of the effective anisotropy that is confirmed by SQUID. The results can be analyzed with the help of measurements of the thermal variation of $K_{eff}$ and $M_s$ at different temperatures (figure 3(c)). As expected for the CoFe transition metal dominant ferrimagnet, lower temperatures lead to a reduction in the saturation magnetization since the rare earth sub-lattice magnetization increase faster than the transition metal one. On the other hand, $K_{eff}$ increases when the temperature decreases. Since the chiral character is kept over the whole temperature range and $|\Psi| = 90°$, eq. 2 can be used to extract the thermal variation of the exchange stiffness of the ferrimagnetic alloy over the whole temperature range. The exchange stiffness is found to be mostly constant with a slight increase when the temperature increases as shown in figure 3(c).

This dependence of the effective exchange on temperature is counterintuitive since in ferromagnetic materials, $A(T)/A(0)$ is expected to vary as $(M(T)/M(0))^\gamma$ with $\gamma$ around 2 [38-39]. Considering that in ferrimagnets the magnon spectrum consists of acoustical and optical branches, Nakamura *et al.* have shown that the temperature dependence of *A(T)* is rather weak up to a certain high temperature [40]. This is due to competing effects of thermal-acoustical and optical magnons, *A(T)* then decreases when T decreases in case of acoustic branch magnons. This prediction has been compared to experimental data by Srivastava *et al.* in case of YIG and magnetite and a satisfactory agreement have been obtained [41]. In their paper, they proposed a comprehensive description by a formula:

$$\frac{A(T)}{A(0)} = \frac{(M_a(T)/M_a(0))(M_b(T)/M_b(0))}{(M_a(T)-M_b(T))/(M_a(0)-M_b(0))} \ (3)$$

where $M_a(T)$ and $M_b(T)$ are the thermal variation of magnetization of the subnetworks a and b, which we now want to use to analyze our data.

While very few experimental data on the variation of $A(T)$ for ferrimagnetic systems are available, and to our knowledge none on the CoFeGd system, we can use experimental data sets acquired by means of element selective X-ray Magnetic Circular Dichroism (XMCD) spectroscopy (see Supplementary S5) a 20 nm thick $Co_{84}Gd_{16}$ alloy (N. Bergeard *et al.*, in preparation) to extract $M_{Co}(T)$ and $M_{Gd}(T)$ and then to derive $A(T)$ using eq. 3. The compensation temperature of this alloy is the same as our 8 nm thick $Gd_{26.1}Fe_{65.5}Co_{8.3}$ alloy. The result of $A(T)$ is shown in figure 3c. We find that $A(T)$ is roughly constant between 80 K and 350 K in line with the experimental data of our GdFeCo ferrimagnet. The good agreement allows us to conclude that the key properties do not vary strongly as a function of temperature showing robust behaviors over the full temperature range.

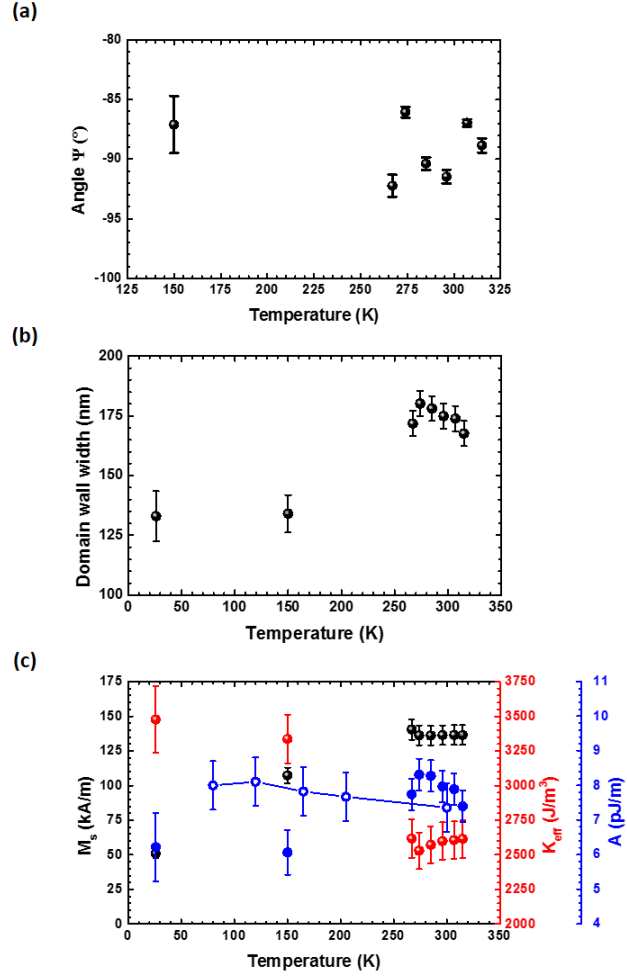

**Figure 3.** Measurements and determinations of magnetic parameters of the ferrimagnet multilayer. (a) Average value of the angle ψ at different temperatures for the TM dominant GdFeCo ferrimagnet. (b) Domain wall width measurements as a function of temperature. (c) Display of the exchange stiffness (blue) for different temperatures as well as the saturation magnetization $M_s$ (black) and the effective anisotropy $K_{eff}$ (red). Blue filled circles: exchange stiffness deduced from domain wall width measurements and magnetic characterizations of our GdFeCo ferrimagnet using eq. 2. The exchange stiffness at 26 K has been calculated assuming an expected pure Néel-type character of the domain wall (|ψ|=90°). Blue open circles: exchange stiffness deduced from XMCD measurements of $Co_{84}Gd_{16}(20)$ using eq. 3 for comparison.

**Conclusion**

In this study, we demonstrate imaging the internal spin structure of domains and domain walls in GdFeCo ferrimagnetic alloys across a 1.5 nm Pt capping layer using SEMPA. This approach is a new path to characterize chiral spin textures in ferrimagnetic multilayers. In the studied GdFeCo-based ferrimagnet, we find that the domain wall spin textures exhibit a pure Néel-type homochirality that is preserved over a large temperature range, even far away from the compensation temperature. This makes GdFeCo a potentially attractive material for skyrmion-based spintronic technologies. Our corrected values of the domain wall width obtained from the high resolution imaging allow us then to extract the exchange stiffness in our material. We can explain the surprisingly temperature dependence of the exchange stiffness from a theoretical model taking into account the multi-sublattices nature of the material. Finally, we report the first direct observation of pure Néel-type skyrmionium in ferrimagnetic materials, quasiparticles that have the advantage of a zero topological charge thus making the material potentially useful for skyrmionic devices.

## Acknowledgment

The authors acknowledge funding from TopDyn, SFB TRR 146, SFB TRR 173 Spin+X (projects A01 & B02). The experimental part of the project was additionally funded by the Deutsche Forschungsgemeinschaft (DFG, German Research Foundation) project No. 403502522 (SPP 2137 Skyrmionics) and the EU (3D MAGIC ERC-2019-SyG 856538. We acknowledge financial support from the Horizon 2020 Framework Programme of the European Commission under FET-Open Grant No. 863155 (s-Nebula). This work was supported by the Institut Carnot ICEEL, by the impact project LUE-N4S, part of the French PIA project "Lorraine Université d'Excellence", reference ANR-15-IDEX-04-LUE, and by the "FEDER-FSE Lorraine et Massif Vosges 2014-2020", a European Union Program. N.K., B.S. and M.K. gratefully acknowledge financial support by the Graduate School of Excellence Materials Science in Mainz (MAINZ, GSC266) and the Max Planck Graduate Center (MPGC). We thank HZB for the allocation of synchrotron radiation beamtime. We acknowledge funding from the French "Agence National de la Recherche" via project No. ANR-11-LABX-0058_NIE, the project EQUIPEX UNION No. ANR-10-EQPX-52 and the CNRS-PICS program.

## Author contributions

The M.K., M.H. and S.M. proposed the study. M.K., M.H., S.M. and R.M.R. supervised the study. B.S, M.H. and J.-L.B grew the samples and B.S and M.H. optimized the samples. D.S., R.M.R., B.S, J.Y. and M.B. performed the SEMPA imaging. B.S, D.S, J.Y. and D.L. performed the analysis of the SEMPA images. B.S., F.K. and N.K. performed the magnetometry. F.R and R.A are the HZB's local contact that operated the ALICE reflectometer while T.K. is in charge of the PM3 beamline at BESSYII. C.B, T.F. and N.B performed the XMCD measurements and took care of the analysis. B.S. drafted the manuscript with the help of M.K., M.H., R.M.R., S.M. and N.B. All the authors commented on the manuscript.

**Competing interests**

The authors declare no competing interests.

# **Supplementary**

**S1: Kerr microscopy at RT**

The sample exhibits a perpendicular magnetic anisotropy (PMA) with a smaller switching field after etching (see figure S1(a)) as expected from the presence of nucleation sites introduced by the Ar ion beam etcher where 3.5 nm of Pt was removed.

In Figure S1(b), we see the spin structures that have been nucleated using a damped oscillating IP magnetic field. We see that micrometer-sized up spin textures have been nucleated that we imaged after with SEMPA to determine the chirality (see main text).

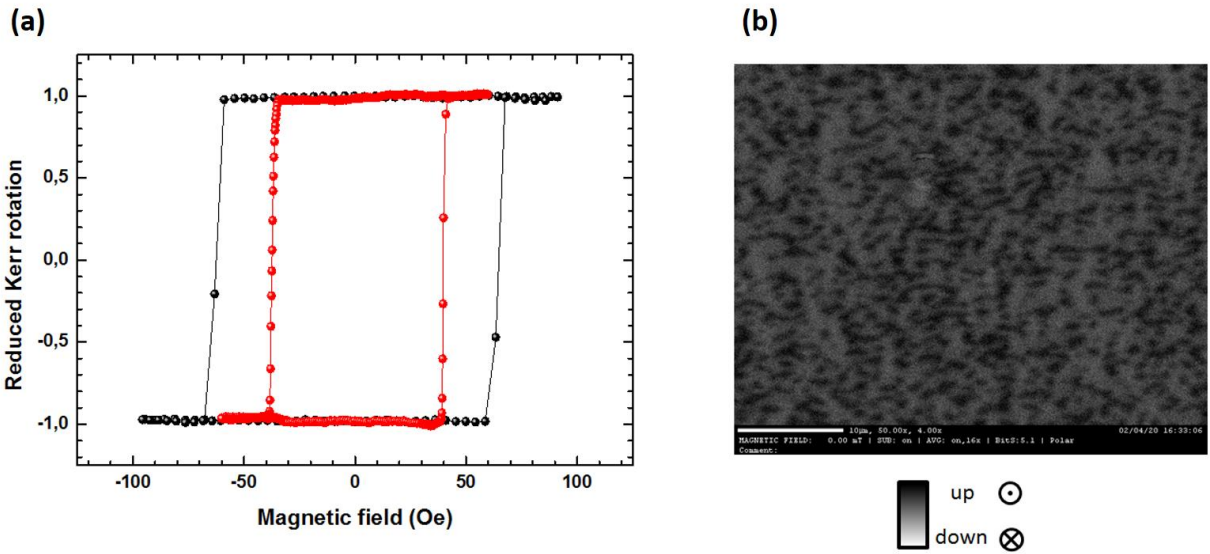

**Figure S1.** (a) Hysteresis loops performed with a polar magneto-optical Kerr effect (MOKE) microscope on Si//Ta(5)/Ir(5)/Fe(0.3)/Gd$_{26.1}$Fe$_{65.5}$Co$_{8.3}$(8)/Pt(5) before (Black) and after milling (red). (b) Image taken with a MOKE microscope at RT after nucleation using an IP magnetic field. Scale bar: 10 μm.

**S2: SEMPA methodology**

To enable surface sensitive SEMPA imaging, Ar ion milling is performed to remove 3.5 nm of the 5 nm Pt capping layer. In SEMPA, a primary beam of 3 nA at 7.5 keV is used to image the DW magnetization orientation. In the case of RE-TM ferrimagnets, the SEMPA acquisition is primarly sensitive to the magnetization of the TM sublattice which then allow to determine the system chirality even in a presence of a reduced net magnetization of the alloy. Since SEMPA is a very surface sensitive technique with a relatively low-efficiency detector [42] and the imaging is performed through a cap layer of 1.5 nm Pt [43], very long acquisition times are needed (about 20 hours per image). The top Pt interface has only been partially removed to retain a sizable negative DMI interaction at the top interface as well as keeping an out-of-plane anisotropy while making the SEMPA imaging possible. Since thermally induced drift of the image during a long acquisition would be detrimental for the required high resolution, we scan the field of view with a fast scan frequency of 1000 pixels/s multiple times subsequently resulting in a total dwell time of 1 ms per pixel. The secondary electrons emitted from the sample surface scatter at the W(001) crystal via low energy electron diffraction (LEED) in different directions depending on their spin. Calculating the normalized asymmetries of the electron flux at the opposite (2,0) LEED spots gives the x and y SEMPA asymmetry images that correspond to the in-plane magnetization of the sample surface [44]. Before further analysis a linear gradient correction is applied to the asymmetry images, which resemble the x and y component of the magnetization vector. From these images, a third image of the absolute magnetization vector is calculated showing the position of the domain walls. Outside the DWs, the measured absolute magnetization value is around zero since in the conventional geometry the system detects the in-plane magnetization component. However, within the DW, a measurable in-plane component exists and is detected. The spatial resolution is determined by imaging the edge of a particle in the SEMPA sum image, where all four detector channels are added up.

**S3: DW width and magnetization angle into the DWs**

From the intensity image of the IP magnetization (see figure S2(a)), we first binarize this image in order to exactly localize the domain walls of each spin textures (see figure S2(b)). In figure S2(c), we shrink the binarized image to get the skeleton of the domain walls, i.e. the one-line at the center of the domain walls. Finally, we cut into several parts the skeleton in order to be able to fit each section by a polynomial function (see figure S2(d)). This will be crucial in order to get the local tangents of the domain walls that will allow us to:

1. Extract the DW width (figure 1(d) in the main text) by measuring and averaging a radial profile at each position of the DWs.

2. Determine the chiral character by comparing the local magnetization direction into the DWs with its local tangent extracted from the polynomial functions (see figure 2(b) in the main text).

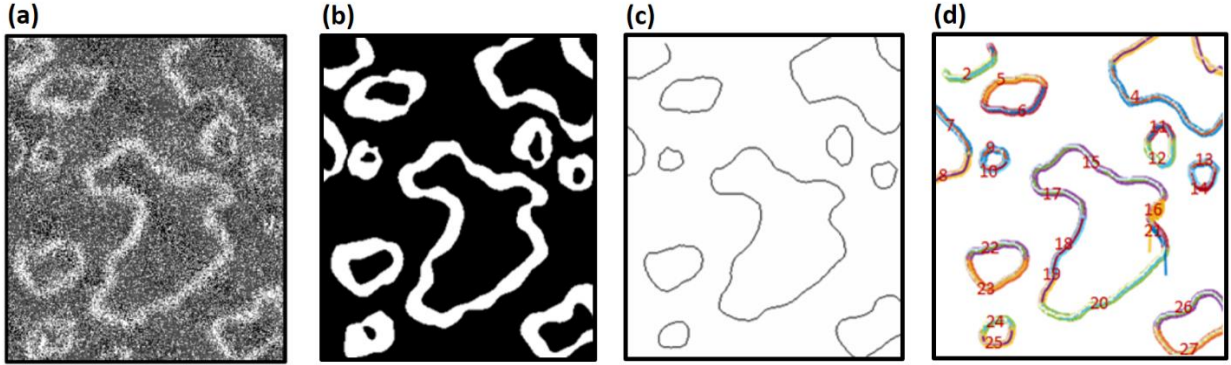

**Figure S2.** (a) IP intensity image taken with SEMPA at RT of our GdFeCo-based ferrimagnetic alloy. (b) Binarization of the image (a). (c) Skeleton of the image (b). (d) Decomposition of the skeleton into several parts allowing a fit of each section of the skeleton with a polynomial function.

## S4: Determination of the SEMPA resolution

The spatial resolution is determined using an edge of a particle in the SEMPA sum image. The edge is assumed to be perfectly sharped even if it maximize our resolution. Therefore the edge defect is modeled as a Heaviside step function. Since the microscope can not be perfectly aligned and that the measured electron beam has a spatial resolution, we consider the imaging technique having an influence on the observed objects that is modeled as a Gaussian function. Finally, the measured edge defect profile is defined as:

$$Edge_{profile} \propto H(x) \otimes e^{-\frac{x^2}{2\sigma^2}}$$

where H(x) represents the Heaviside step function and the Gaussian function represents the influence of the imaging technique on the real edge defect (Figure S3a). On the Figure 1b, we fit the edge profile with our convolution function. The resolution is commonly measured as the spatial interval corresponding to a variation of the intensity between 20% and 80% of a sharp edge [45]. We finally found that the resolution is about 28 nm at RT.

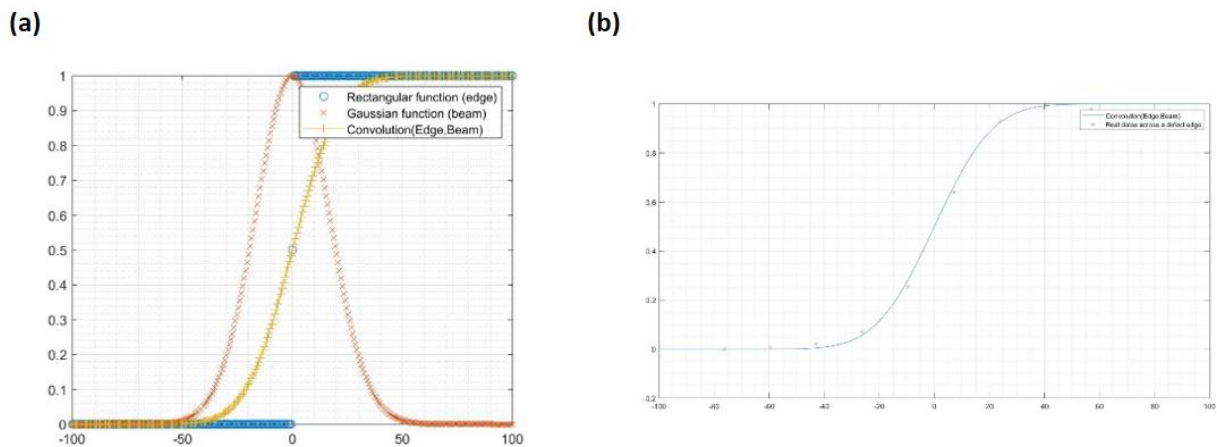

**Figure S3.** x: distance in nm. y: reduced intensity between 0 and 1. (a) Simulated edge profile (yellow) calculated from the convolution between a Heaviside step function (blue) with a Gaussian function (orange) (see eq. 1 in the main text) using arbitrary parameters. (b) Real edge profile measured across a defect on a SEMPA image (red crosses) with the fitting edge profile function.

**S5: XMCD spectroscopy methodology**

The measurements on CoGd have been performed by using the ALICE reflectometer installed on the PM3 beamline at the BESSY II synchrotron radiation source operated by the Helmholtz-Zentrum Berlin [46]. The X-ray Absorption Spectra (XAS) at the Co L3 and Gd M5 edges were acquired by monitoring the transmission of circularly polarized X-ray under a magnetic field of +/-1 kOe. The magnetic field was applied along the X-ray propagation while the sample was tilted by 30° according to the in-plane magnetic anisotropy of the alloys.

**Supplementary references**